\documentclass[sigconf, nonacm=true]{acmart}
\makeatletter
\renewcommand\@formatdoi[1]{\ignorespaces}
\makeatother
\AtBeginDocument{%
  \providecommand\BibTeX{{%
    \normalfont B\kern-0.5em{\scshape i\kern-0.25em b}\kern-0.8em\TeX}}}
    
\newcommand\blfootnote[1]{%
\begingroup
\renewcommand\thefootnote{}\footnote{#1}%
\addtocounter{footnote}{-1}%
\endgroup
}

\setcopyright{none}
\renewcommand\footnotetextcopyrightpermission[1]{} 
\begin{document}

\title{Enabling Heterogeneous, Multicore SoC Research\\ with RISC-V and ESP}

\author{Joseph Zuckerman, Paolo Mantovani\footnotemark, Davide Giri, and Luca P. Carloni}
\email{{jzuck, paolo, davide\_giri, luca}@cs.columbia.edu}
\affiliation{%
  \department{Department of Computer Science}
  \institution{Columbia University}
  \city{New York}
  \state{New York}
  \country{USA}
  }

\renewcommand{\shortauthors}{J. Zuckerman, P. Mantovani, D. Giri, and L.P. Carloni}


\begin{abstract}
    Heterogeneous, multicore SoC architectures are a critical component of today's computing landscape. However, supporting both increasing heterogeneity and multicore execution are significant design challenges. Meanwhile, the growing RISC-V and open-source hardware (OSH) movements have resulted in an increased number of open-source RISC-V processor implementations; however, there are fewer open-source SoC design platforms that integrate these processor cores. We present modifications to ESP, an open-source SoC design platform, to enable multicore execution with the RISC-V CVA6 processor. Our implementation is modular and based on standardized interfaces. These properties simplify the integration of new cores. Our modifications enable RISC-V-based SoCs designed with ESP for FPGA to boot Linux SMP and execute multithreaded applications. Coupled with ESP's emphasis on accelerator-centric architectures, our contributions enable the seamless design of a wide range of heterogeneous, multicore SoCs.   
\end{abstract}
\settopmatter{printfolios=true}
\settopmatter{printacmref=false}

\maketitle

\section{Introduction}
\blfootnote{*Paolo Mantovani is now with Google Research.}
Modern computing systems increasingly rely on heterogeneous system-on-chip (SoC) architectures, which combine general-purpose processors with specialized hardware accelerators \cite{cong_dac14, xavier_hc18, xilinx_hc18}. This shift toward heterogeneity, however, comes with new challenges, as the required engineering effort scales with increasing heterogeneity \cite{khailany_dac18}. The open source hardware (OSH) movement, largely enabled by the RISC-V community, can mitigate these rising non-recurring engineering costs by promoting collaborative engineering and design reuse \cite{gupta_osh17, asanovic_riscv}.  

The diminishing returns of performance scaling by exploiting parallelism with multicore architectures is one of the primary reasons for the rise of heterogeneous architectures. Still, featuring multiple coherent processor cores remains a key property for most SoCs. Utilizing multiple cores, however, adds significant complexity to SoC design because maintaining coherence and enforcing synchronization among cores are challenging tasks. To enable future generations of heterogeneous, multicore SoC architectures, we propose a platform that allows for seamless design of multicore SoCs containing heterogeneous intellectual property (IP) blocks.

This paper presents modifications to ESP, an open-source platform for heterogeneous SoC design, to enable SoCs with up to 4 RISC-V CVA6 cores that can boot Linux SMP and execute multithreaded applications. ESP is an open-source research platform for heterogeneous SoC design that combines a scalable tile-based architecture and a flexible system-level design methodology \cite{mantovani_iccad20}. The architecture simplifies the integration of heterogeneous components developed by different teams, and the methodology embraces the use of various design flows for component development \cite{carloni_dac16}. ESP enables rapid FPGA-based prototyping of SoC architectures \cite{mantovani_dac16} and provides the foundation for the agile design, optimization, and tapeout of SoCs \cite{jia_esscirc2022}. Prior to this work, ESP supported multicore SPARC-based SoCs and single core RISC-V-based SoCs. The support for multiple RISC-V cores was implemented in a standardized manner that can simplify the integration of new cores that are AXI-compliant. When this is coupled with ESP's scalable architecture, agile methodology, and multiple design flows for hardware accelerators, ESP can drastically increase design productivity for a wide range of heterogeneous, multicore SoCs. All of the contributions described in this paper are included in the open-source release of ESP \cite{esp}. 

The rest of the paper is organized as follows. We first give some background on ESP (Section 2), and then highlight the changes required to enable multiple RISC-V cores (Section 3). Then, we present some experimental results from running a multithreaded benchmark suite on top of ESP (Section 4). Finally, we discuss the differences between ESP and other multicore RISC-V platforms (Section 5) and conclude (Section 6). 

\blfootnote{To appear in the Sixth Workshop on Computer Architecture Research with RISC-V (CARRV 2022)}

\begin{figure}
    \centering
    \includegraphics[width=\linewidth]{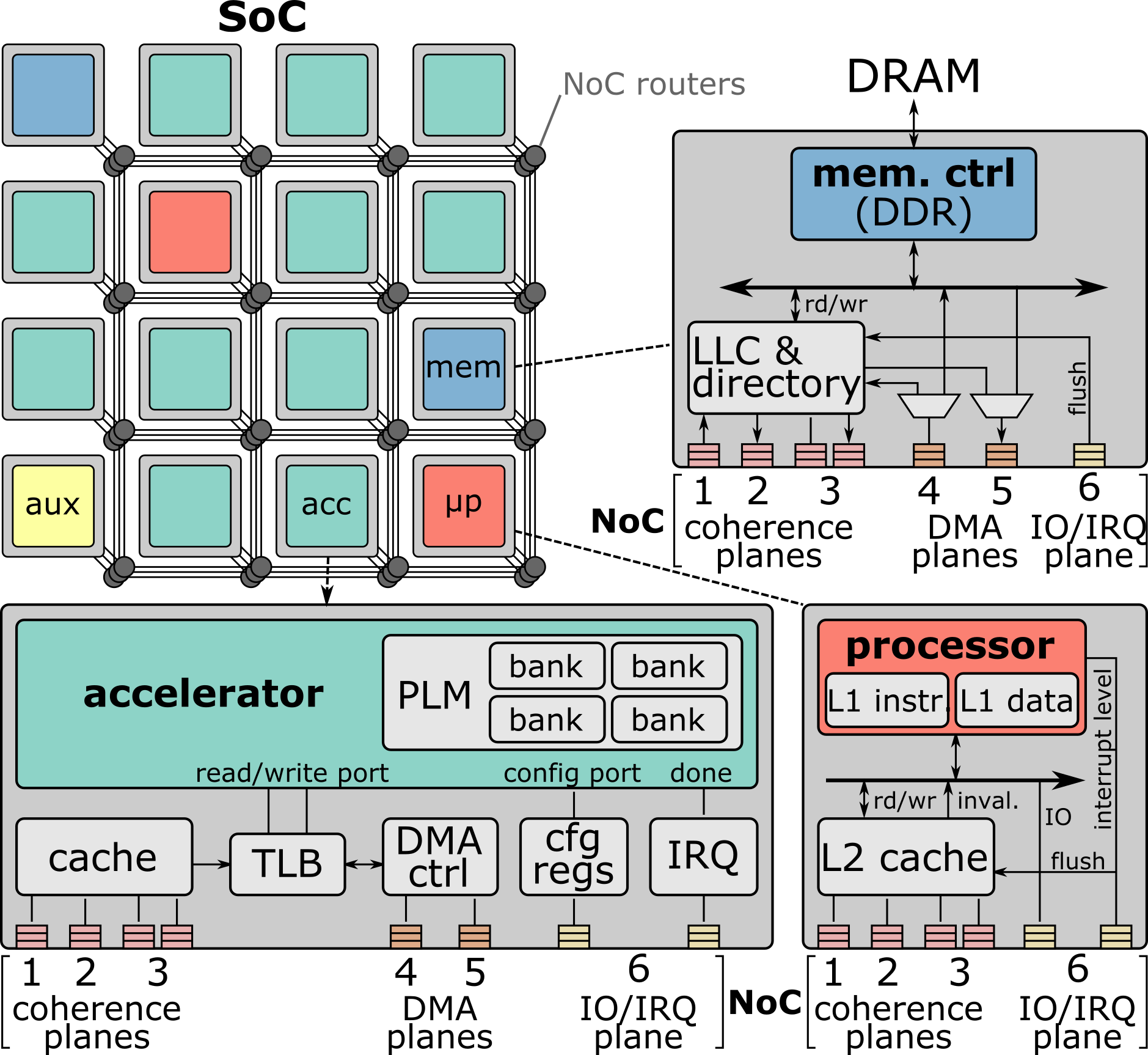}
    \caption{A 4x4 ESP tile grid with 2 processor, 2 memory, 11 accelerator, and 1 auxiliary tile. The figure shows how the various tiles interface to the 6 physical NoC planes.}
    \label{fig:esp}
\end{figure}

\section{The ESP Architecture}

The ESP architecture is structured as a heterogeneous tile grid, as shown in Figure \ref{fig:esp} \cite{mantovani_iccad20, carloni_dac16}.  At SoC design time, users can specify an ESP instance by selecting the dimensions of the grid and the type of each tile with assistance from the ESP graphical user interface.  To facilitate scalability, the tiles are interconnected by a multi-plane network-on-chip (NoC). A key feature of the ESP architecture is the modular \textit{socket}, which interfaces each tile to the NoC. The socket decouples the design of the tile from the rest of the SoC and greatly simplifies IP integration, following the principles of communication-based system-level design \cite{carloni_2015}. The socket also implements several \textit{platform services}, such as dynamic voltage frequency scaling (DVFS) and performance monitors, that ``come for free'' with the addition of a new IP. There are 4 main types of tiles.

\textbf{Processor Tile.} Each processor tile contains a CPU from one of several options, including the 32-bit SPARC Leon3 core \cite{leon3}, the 64-bit RISC-V CVA6 (formerly known as Ariane) \cite{ariane_paper}, and the 32-bit RISC-V Ibex (formerly known as Zero-riscy). Only one processor type can be used per SoC, and prior to this work, an SoC could instantiate multiple processor tiles only if the Leon3 core was selected. The CPU in the processor tile serves as the host of the system and can boot an operating system in the case of Leon3 and CVA6. The processors are instantiated with their own private L1 cache. If the ESP cache hierarchy is enabled -- as required for multicore SoCs -- the processor tile also instantiates an ESP L2 cache. The L2 cache allows the CPU to transparently participate in the ESP coherence protocol, which is described in Section 2.1. 

\textbf{Memory Tile.} Each memory tile instantiates a channel to external DRAM. The memory tile also includes a slice of the last-level cache when the cache hierarchy is enabled. To improve off-chip bandwidth, multiple memory tiles can be instantiated. When multiple memory tiles are used, the global address space is split, and each memory tile services a discrete partition of the address space. All logic to handle the partitioned address space is automatically generated and is completely transparent to software.

\textbf{Accelerator Tile.} The accelerator tile instantiates a loosely-coupled accelerator, which executes a coarse-grained task \cite{cota_dac15}. There are two main types of accelerators: accelerators designed with ESP and accelerators designed by a third-party. For accelerators designed with ESP, the socket provides several services that lower the design effort, such as configuration registers, DMA, and virtual memory. ESP supports several different flows for accelerator design, which feature a high degree of automation in all cases. The main classes of design flows are 1) Verilog/VHDL, 2) C, C++, or SystemC with Stratus, Catapult, or Vivado HLS, and 3) directly from machine-learning models in Keras, TensorFlow, PyTorch, or ONNX with HLS4ML \cite{giri_date20}. For accelerators that are designed independently of ESP, a third-party flow exists to seamlessly integrate accelerators that comply with a standard bus protocol, like AXI, such as the NVIDIA Deep Learning Accelerator (NVDLA) \cite{giri_ieeemicro21, nvdla}. 

An invocation of a loosely-coupled accelerator begins with configuration from software (using a device driver if an OS is running) via memory-mapped registers. After it is started, the accelerator issues DMA requests, which are sent to the ESP memory hierarchy using one of four cache-coherence modes. The different cache-coherence modes operate completely transparently to the accelerator and can be selected at runtime based on the needs of the accelerator and the overall status of the system \cite{giri_aspdac19, zuckerman_micro21}. When the accelerator completes, it sends an interrupt that resumes the execution of the software thread that invoked it.  

\textbf{Auxiliary Tile}. The auxiliary tile handles non-memory I/O, like Ethernet and UART. It also includes several miscellaneous components like the interrupt controller, boot ROM, and frame buffer. The Ethernet connection both enables remote connection via SSH and the \textit{ESPLink} debug application, which allows a host machine to access any memory-mapped regions of the SoC and load program binaries into on-chip and off-chip memory. 

\textbf{Network-on-Chip.} The ESP NoC has a packet-switched 2D-mesh architecture with lookahead routing. Each tile contains a single NoC router, and every hop between adjacent tiles takes a single cycle. The NoC comprises six physical planes to provide ample bandwidth and prevent deadlock. Three planes are for coherence messages, two for DMA, and one for memory-mapped register access and interrupts. 

Each tile instantiates a local bus, which enables communication between components of the same tile. ESP uses a set of \textit{proxies} that convert bus requests to NoC messages, and vice versa, for communication between components in different tiles. Thanks to the proxies, remote components can issue transactions to each other as if they were connected to the same bus. 

\subsection{The ESP Coherence Protocol}

ESP's coherence protocol is an extension of a standard MESI directory-based protocol \cite{nagarajan_slca20}. The protocol is adapted to both work over a NoC and support last-level cache coherent accelerators \cite{giri_nocs18}. The L2 and LLC caches implement the coherence protocol and were originally designed to support multicore execution for the SPARC Leon3 core.

\textbf{Last-Level Cache}. The LLC sits in the memory tile and is the final layer of on-chip storage before DRAM. The LLC includes the directory controller, which stores metadata about every cache line. In addition to the four standard MESI states, the LLC adds a Valid (V) state. The Valid state indicates that there are neither any active sharers nor an owner of the cache line, but the data in the cache line is valid. Accessing a cache line in the Valid state does not require a request to main memory, which would normally occur from the Invalid state. 

In addition to the standard coherence messages (request, response, forward), the LLC can also receive DMA transactions to support LLC-coherent accelerators. DMA bursts are accepted as one cache request, but are handled on a cache-line granularity. After accelerators read and write data, the cache lines they access are left in the Valid state. To ensure functional correctness, accelerators must have mutual exclusivity over their data while they operate. While the LLC-coherent mode offers benefits, sometimes a non-coherent access mode is more appropriate, particularly in the case of large datasets that don't fit on chip \cite{giri_ieeemicro18}. Before a non-coherent accelerator execution, the last-level cache must be flushed to ensure all processor writes are reflected in main memory. The accelerator's device driver actuates the flush of the LLC by writing to a memory-mapped register in the socket of the memory tile (or to multiple such registers when the LLC is distributed across multiple memory tiles). 

\textbf{L2 Cache.} The L2 cache is a private cache that can be instantiated in processor and accelerator tiles. When an accelerator is equipped with a private cache, it can use the fully-coherent access mode, in which it participates in the system's coherence protocol just like a processor core. Due to race conditions that can arise over the different NoC planes, several transient states are needed to guarantee correct operation. A small set of miss-status handling registers (MSHR) keep track of the status of ongoing transactions. Incoming requests and forwards messages can be stalled if they conflict with an ongoing transaction, depending on the state of the corresponding cache line.

Supporting invalidation of the L1 cache and atomic instructions are two key features of the L2 cache that enable multicore execution. The Leon3 core sits on an AHB bus, which also interfaces with the L2 cache. To invalidate the L1, a fake write is issued on the AHB bus from the L2. The Leon3 core, which supports snooping on the AHB bus, sees the write and invalidates the corresponding entry in its L1. For the SPARC instruction set, the L2 handles the test-and-set and compare-and-swap atomic instructions. For these instructions, the processor issues one or more loads to a cache line, potentially followed by one store to the same cache line. A special transient state is used to indicate that a cache line has been accessed by an atomic read, but not by the atomic write that completes the instruction. During this period, the core executing the atomic instruction is the owner of this cache line, and any forward requests for the same cache line are stalled to guarantee atomicity. The next section will discuss how L1 invalidation and atomic instructions are implemented for the RISC-V CVA6 core. These were the two key challenges that we had to address in order to support multicore execution.

\section{Enabling Multicore RISC-V}

\begin{figure}
    \centering
    \includegraphics[width=\linewidth]{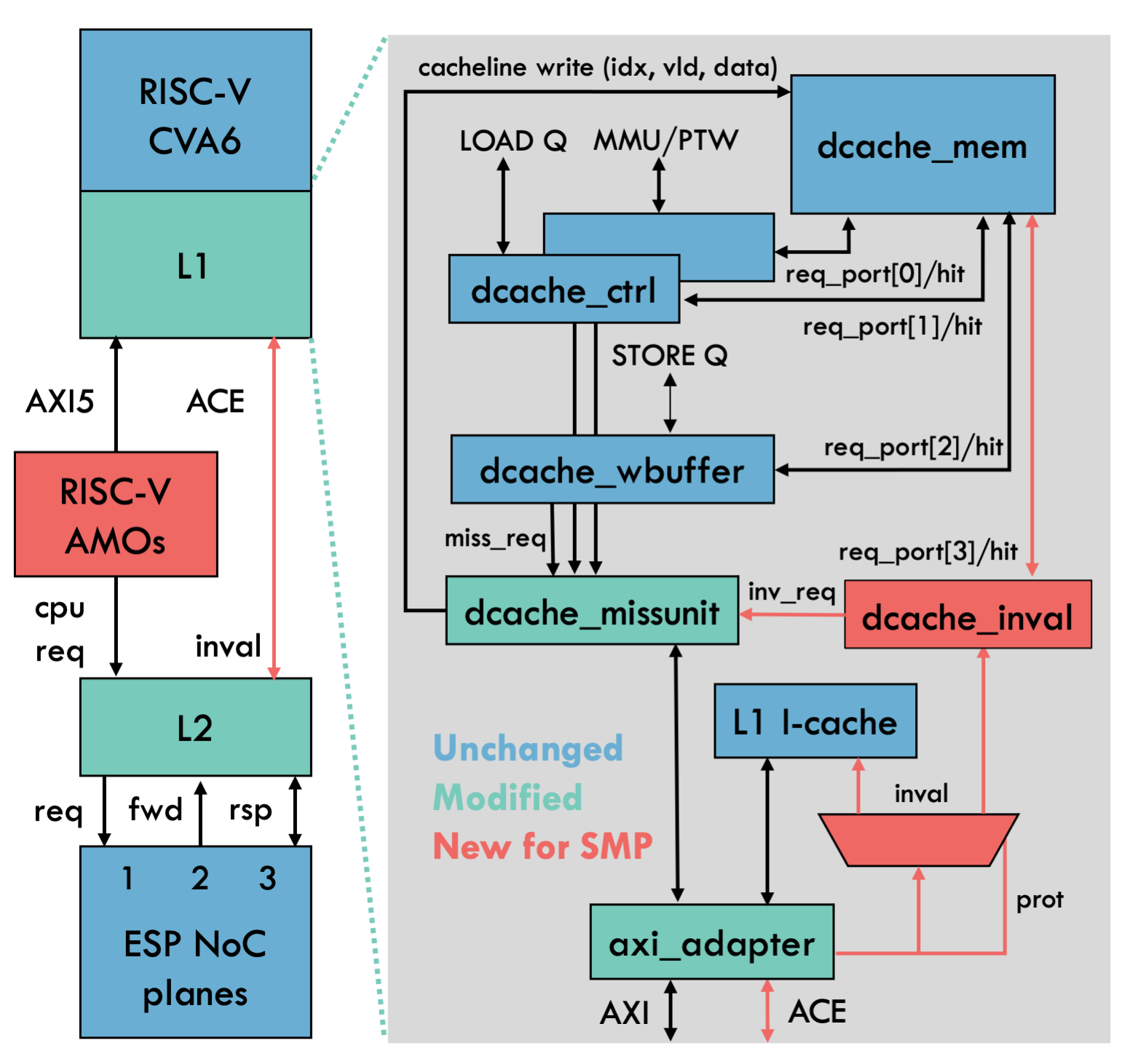}
    \caption{Modifications to the ESP processor tile and CVA6 L1 cache to enable multicore execution.}
    \label{fig:l1}
\end{figure}

Prior work, which first integrated the CVA6 (at the time, Ariane) core with ESP, only enabled single core execution \cite{giri_carrv20}. As shown in Figure \ref{fig:l1} , the main changes to ESP to support multicore execution with the RISC-V CVA6 core include 1) modifying the core to accept invalidations of its L1 cache using the AXI Coherency Extensions, and 2) supporting RISC-V atomic and LR/SC instructions with changes to the L2 and AXI interconnect. 

Since this work revolves around the ESP cache hierarchy, it is worth briefly discussing how prior work integrated the CVA6 core with the ESP caches. With the SPARC Leon3 core, the processor tile instantiates an AHB bus, to which the processor attaches. A wrapper around the L2 cache converts AHB transactions to cache requests, and vice versa, for cache responses. Since the CVA6 core uses an AXI interface, the processor tile instantiates an AXI crossbar instead of the AHB bus when the CVA6 core is selected. Hence, a new L2 wrapper was developed to interface the cache with the AXI protocol. The only other change to the L2 cache was modifying the write logic to support both little-endian and big-endian architectures, since CVA6 is little-endian, while Leon3 is big-endian. The last-level cache did not change at all, since it is completely decoupled from the implementation of the processor by the L2. 

\subsection{L1 Invalidation of CVA6}
The CVA6 is not natively multicore, as it does not accept invalidations of its L1 through the primary AXI interface. Prior work, however, integrated CVA6 into the OpenPiton architecture to enable multicore execution \cite{balkind_asplos16, balkind_carrv19}. CVA6 connects to OpenPiton's L1.5 cache through a custom interface. The L1 accepts invalidations from the L1.5 by receiving both the address and way in the L1 set to invalidate. This means that the L1.5 must store the corresponding L1 way of each cache line to send invalidations. Because of this extra area overhead and to support a more standard interface, we choose to use the AXI Coherency Extensions (ACE) to send invalidations from the ESP L2 and modify the CVA6 core in a minimal manner to receive them in this fashion.

ACE adds three channels -- Snoop Address (AC), Snoop Response (CR), and Snoop Data (CD) -- to the original five of AXI to enable coherence between different components in broader systems \cite{arm_ace}. Because invalidation requires neither responses nor data, we only instantiate the AC channel in the processor tile and modify the interface of the core to receive messages on this channel. Upon an invalidation, the L2 cache drives the snoop address bus with a \texttt{MakeInvalid} transaction and the address of the line to invalidate. The L2 also sends the permission bits of the cache line to the core in order to easily route the invalidations to either the L1 data cache or L1 instruction cache; this was previously not necessary for the Leon3 core. 

To execute the invalidation, we make slight modifications to only the L1 cache of CVA6; the core's pipeline remains unchanged. We create a new data cache invalidate unit 
(\texttt{wt\_dcache\_inval}) that snoops for \texttt{MakeInvalid} commands on the AC channel. Upon receiving an invalidation, the invalidate unit performs a lookup in the data cache memory through a new dedicated port to check for the requested address. If the address lookup results in a hit, then the cache line is invalidated; else, the invalidate request is ignored. While we currently route instruction invalidations to the L1 instruction cache, we choose not to modify the instruction cache to support invalidations, since instruction memory is currently never modified in ESP. Since the instruction cache's memory only has one port, enabling its invalidation would require modifying the instruction cache's state machine, as opposed to the less-invasive addition of a new unit for the data cache. However, the state machine is relatively simple, and the required changes would not be extensive.

When an ESP accelerator runs in a LLC-coherent or non-coherent manner, the L2 is flushed via a memory-mapped register write from software. When this happens, the entire L1 must be invalidated or flushed. In the case of the Leon3 core, a SPARC flush instruction is used to trigger an L1 flush before signaling for the L2 to flush. Since the RISC-V instruction set supported by the CVA6 core does not have a flush instruction, we choose to expose the CVA6's L1 flush signal at the interface of the core. The same signal that triggers the L2 flush is connected to this port of the core, and the L2 waits for the L1 to finish flushing using an additional \texttt{flush\_done} signal at the interface. It would be possible to use the new ACE bus to invalidate one line of the L1 at a time from the L2 and avoid additional modifications to the core's interface, but this would be far less efficient.

\subsection{Handling RISC-V Atomics and LR/SC}

RISC-V atomic memory operations (AMOs) are handled by instantiating an AXI Adapter for RISC-V Atomics from the Pulp Platform \cite{axi_riscv_atomics}. The adapter is placed on the AXI bus between the L1 and L2 caches. AMO requests exit the core as a single transaction on the address write (AW) channel. The adapter converts this AMO request into separate read and write transactions on the downstream AR and AW channels, respectively. After the adapter receives the read response from the L2, a small ALU performs the computation indicated by the instruction, and it issues the write to the L2 with the result. We make a slight modification to the adapter to set the \texttt{lock} field on both channels to signal to the L2 that the read and write are part of an atomic operation. 

The atomicity of AMOs is enforced by the L2. After receiving an atomic read request, the L2 issues a \texttt{GetModified} request to the LLC if it does not already have the corresponding cache line in a modified state. The line then transitions to the transient \texttt{XMW} state, indicating that an atomic operation has started and the write is pending. While in this state, the L2 will stall all forward requests to this line, such that no other core can read or write to this line during the atomic operation. All of this logic leverages the prior implementation of SPARC atomics for the Leon3 core. However, because RISC-V atomics contain exactly one read and one write, while SPARC atomics consist of one or more reads potentially followed by a write, some minor modifications are needed to appropriately determine when AMOs are completed. 

We utilize the infrastructure for handling AMOs to also handle the RISC-V load-reserved (LR) and store-conditional (SC) instructions, but with several modifications. First, because the SC instruction may not succeed, we modify the L2 to provide the write response (\texttt{bresp}) on the AXI bus; the code is \texttt{EXOKAY} in case of success and \texttt{OKAY} in the case of failure. The AXI RISC-V AMOs adapter is also slightly modified to forward this response to the core, instead of replying with \texttt{EXOKAY} by default. Second, forward requests to the same cache line are served between LR and SC operations. However, after they are served, the atomic operation is marked as completed. Then, if the SC operation arrives at the L2, the write does not take place, and the cache replies with \texttt{OKAY}, designating failure. Finally, for SPARC atomics, any non-atomic read or write that follows a read atomic is assumed to end the ongoing atomic operation. This is also the case for any reads or writes to data that follow an LR for RISC-V, because the ISA prohibits loads and stores between LR and SC instructions. However, instruction fetches must be served between the LR and SC pairs. This requires the most significant modifications to the L2, which was not previously designed to handle any memory accesses during ongoing atomic operations. Certain cases that previously triggered stalls, such as an instruction fetch that maps the same cache set as an ongoing atomic, must be overridden. 

The \texttt{lock} field on the AXI channels is also used to signify the LR and SC operations as atomic. To distinguish between an LR/SC pair and an AMO, we use the \texttt{atop} field, which is zero for LR/SC and non-zero for AMOs. However, since the read arrives first and there is no \texttt{atop} field on the AR bus, we make use of the \texttt{user} field to send this information. 

\section{Experimental Results}

\begin{figure}
    \centering
    \includegraphics[width=0.65\linewidth]{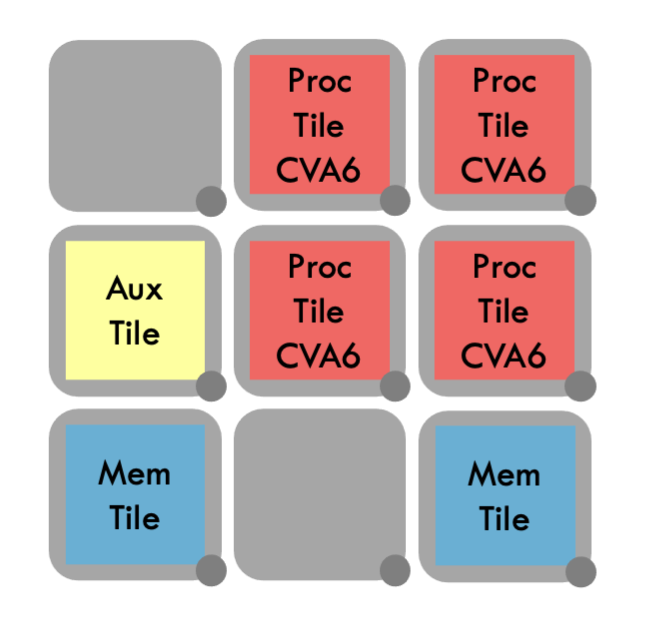}
    \caption{The evaluation SoC with 4 CVA6 processor tiles, 2 memory tiles, and the auxiliary tile.}
    \label{fig:socs}
\end{figure}

\begin{figure*}
    \centering
    \includegraphics[width=\textwidth]{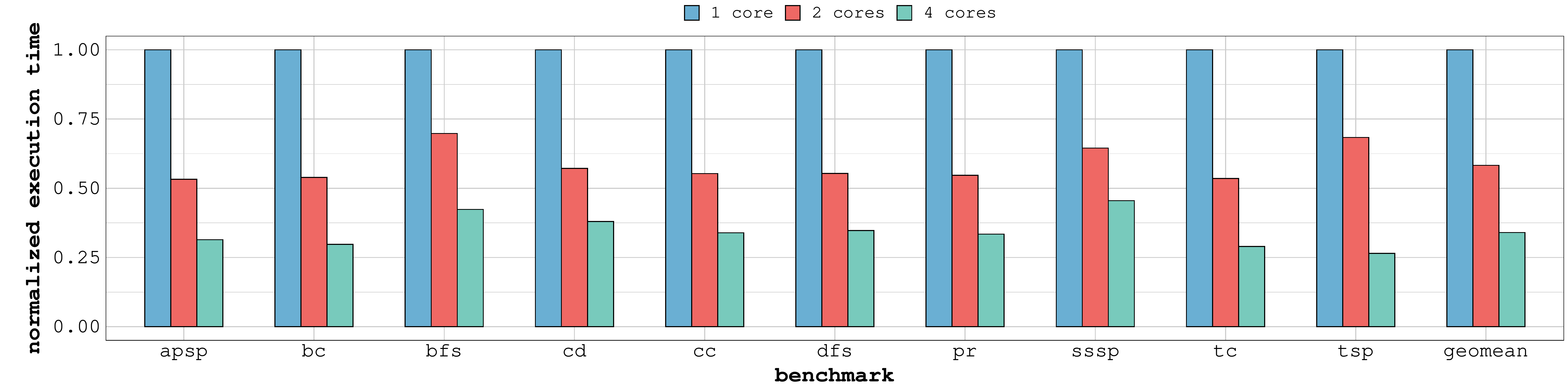}
    \caption{Execution time of the CRONO benchmark suite on 1, 2, and 4 CVA6 cores in ESP. }
    \label{fig:crono}
\end{figure*}

Implementing the changes for the CVA6 invalidation and RISC-V atomic instructions in ESP took a few weeks each. Shortly thereafter, we ran the first multicore RISC-V baremetal programs in ESP. In a matter of days from this point, we also booted Linux SMP on 4 RISC-V cores for the first time. Roughly an additional 2 months were spent debugging minor issues to be able to consistently boot Linux SMP and run intensive multithreaded applications on 4 CVA6 cores. 

To stress test our implementation and also evaluate its performance, we use the CRONO benchmark suite for multithreaded graph algorithms \cite{crono}, which contains the following algorithms: \\
\textbullet\ \textit{Path planning}: Single Source Shortest Path (SSSP), All Pairs Shortest Path (APSP), Betweenness Centrality (BC). \\
\textbullet\ \textit{Search}: Breadth First Search (BFS), Depth First Search (DFS), Traveling Salesman Problem (TSP).\\
\textbullet\ \textit{Graph Processing}: Connected Components (CC), Triangle Counting (TC), PageRank (PR), Community Detection (CD).\\

The CRONO suite was useful in revealing the (hopefully) last few implementation bugs. The successful execution of all of its applications gives us further confidence in our design.


Using the ESP SoC generation GUI, we prepared an evaluation SoC, as shown in Figure \ref{fig:socs}. The SoC contains 4 instances of the CVA6 processor, 2 memory tiles, and the auxiliary tile. Each memory tile has a 512KB LLC slice, for a 1MB aggregate LLC. The L2 cache in each processor tile is 64KB.  We note that this is just one possible ESP SoC with 4 CVA6 cores, and the SoC could be seamlessly altered to change any of the following parameters: number of memory tiles, presence of accelerator(s), NoC dimensions, cache sizes, and positions of each tile. This SoC is generated for a proFPGA quad Virtex Ultrascale Prototyping System, which mounts Xilinx XCVU440 FPGAs. ESP runs at 78 MHz on this FPGA.

The applications run on top of Linux SMP with a configurable number of threads. We evaluate all of the benchmarks on 1, 2, and 4 cores, using the same number of threads as cores. We make no modifications to the implementation of the algorithms and cross compile them for 64-bit RISC-V.


Figure \ref{fig:crono} shows the execution time of each benchmark on 1, 2, and 4 cores, normalized to the performance of 1 core. For 2 cores, the execution time ranges from 53\% to 70\% of the single core performance, with a geometric mean of 58\%. For 4 cores, the range is from 26\% to 42\%, with a geometric mean of 34\%. These figures roughly match the performance of the CRONO benchmark performed by its authors on a multicore simulator \cite{crono}. These results confirm the scalability of the ESP architecture for running multithreaded applications on multiple processors. 
\section{Related Work}

Here, we discuss a few other open-source SoC platforms that offer multicore RISC-V execution, and then discuss the key distinguishing features of ESP. 

\textbf{OpenPiton} is a manycore research framework, designed to enable academic research of manycore systems \cite{balkind_asplos16}. Like ESP, OpenPiton has a tile-based architecture on top of a multi-plane NoC. OpenPiton was originally developed with support for the Open\-SPARC T1 core. Each OpenPiton tile contains a private L1.5 cache and a slice of the distributed L2 cache. A \textit{chipset} in an OpenPiton chip provides access to external DRAM and other I/O, as well as a connection to other OpenPiton chips; the chipset is decoupled from the NoC via a \textit{chip bridge}. 

More recently, OpenPiton has also added support for the RISC-V CVA6 (64-bit) and PicoRV32 (32-bit) cores, as well as the x86 \texttt{ao486} core \cite{balkind_byoc}. These cores, along with the OpenSPARC T1, can coexist in the same OpenPiton system, thanks to a new Transaction-Response Interface (TRI). As previously mentioned, the CVA6 connects to OpenPiton's L1.5 through a custom interface. The L1.5 sends invalidations to the L1 by sending the correct cache way to invalidate, which requires extra storage in the L1.5. The PicoRV32 core does have an L1 cache and is not Linux capable, while invalidation of the \texttt{ao486} was still underway at the time of the writing of \cite{balkind_byoc}; hence, both cores must be used in a singlecore mode. OpenPiton also supports the NVIDIA Deep Learning Accelerator and the MIAOW GPU by connecting these accelerators to the crossbar in the chipset.

\textbf{Black Parrot} is a Linux-capable RISC-V multicore, which aims to be a host for accelerator SoCs \cite{blackparrot}. The BlackParrot system is built around the BlackParrot core, which aims to be ``Tiny, Modular, and Friendly'', and achieves impressive CoreMark scores compared to other academic and industry RISC-V processors \cite{blackparrot}. The BlackParrot system architecture is also tile-based with multiple NoCs. Coherence is enforced with the BedRock implementation, which consists of local-cache engines (LCEs) connected to the private cache of processing elements and cache-coherence engines (CCEs), which collectively implement the system's coherence directory and controller. BlackParrot supports both coherent accelerators, which utilize an LCE, and streaming accelerators, which do not have a backing cache. BlackParrot does not rely on standard bus protocols, like AXI and AHB, but does provide a set of adapters for these protocols. 

\textbf{Chipyard} is an SoC-generation framework that combines several projects for agile hardware design \cite{chipyard}. Chipyard is primarily built around Rocket Chip, a Chisel-based SoC generator \cite{rocket}. Rocket Chip generates RTL for a full SoC, including the processor cores, tightly-coupled accelerators (Rocket Custom Coprocessor or ROCC), loosely-coupled accelerators (MMIO accelerators), peripherals, and the system interconnect, which implements the TileLink protocol \cite{tilelink}. Chipyard adds additional components to the IP library, including the CVA6 core, out-of-order cores (BOOM), vector processors, and deep learning accelerators. Chipyard supports FPGA-accelerated cycle-accurate simulation through the FireSim platform \cite{firesim} and also has an agile VLSI flow (Hammer). Centrifuge \cite{centrifuge} leverages FireSim to evaluate SoCs that utilize HLS-generated accelerators.

\textbf{HERO} combines a Linux-capable host with a programmable manycore accelerator (PMCA). The first version of HERO \cite{kurth_hero} uses an ARM Cortex-A multicore processor as the host; the PMCA consists of clusters of RISC-V PEs with shared memory that connect to the host via the system interconnect. Hence, the entire system uses multiple ISAs. HEROv2 \cite{kurth_herov2} improves upon the first version by enabling 64-bit hosts to coexist with the the 32-bit PEs in the PMCA. HEROv2 also adds the CVA6 core as a host option, but only supports a single core option in this case. HEROv2 features a complete heterogeneous compiler based on LLVM that can support OpenMP applications.

\textbf{MEG} is a full system emulation infrastructure for evaluating software and hardware designs in systems utilizing high-bandwidth memory (HBM) \cite{meg}. MEG uses the BOOM core and TileLink interconnect generated from Chipyard to create a Linux-capable multicore infrastructure on FPGA. MEG also facilitates the hardware and software integration of AXI-compliant hardware accelerators, which connect to the IOMMU in the system. A custom performance monitoring system allows studying the impacts of HBM on various aspects of the system. 

\textbf{ESP} has the following three key characteristics, which, together, distinguish it from the other mentioned platforms:\\
1) By using \textit{standardized bus protocols and adapters}, ESP's architecture is decoupled from any particular core. Furthermore, this simplifies the integration of new cores that utilize the same standardized protocols. For example, the T-head (founded by Alibaba) Xuantie C910 core makes use of AXI and the AXI Coherency Extensions \cite{xuantiec910}.\\
2) ESP is designed with a \textit{system-centric mindset} rather than a processor-centric one. Loosely-coupled accelerators are given an equal importance in the SoC architecture as processor cores. They occupy their own tile connected to the NoC, rather than being attached to the I/O bus. Furthermore, ESP provides several design flows to create new, custom hardware accelerators, which can work seamlessly with any processor core that is integrated in ESP. \\
3) ESP's \textit{scalable architecture} allows the design of large SoCs. The NoC can support large grids of processing elements that would not be feasible in bus- or crossbar-based systems. Furthermore, the distributed memory hierarchy can support multiple channels to off-chip memory in order to satisfy increased bandwidth requirements of large designs. 

\section{Conclusion}
We augmented the open-source ESP project to enable multicore execution
with the RISC-V CVA6 processor core.
In doing so, we provided the CVA6 processor with mechanisms to invalidate its L1 cache and enabled the ESP cache hierarchy to support RISC-V atomic instructions. We implemented these mechanisms in a modular way by relying on standard bus protocols. 
Hence, our implementation could be reused to simplify the integration of other processor cores that are AXI-compliant. We verified the correctness and performance of our implementation by booting Linux SMP and running multithreaded graph applications on an SoC with 4 CVA6 cores on FPGA. All of this work is included in the open-source release of ESP.

\begin{acks}
This work was supported in part by DARPA (C\#:HR001118C0122),
the ARO (G\#:W911NF-19-1-0476), and the NSF Graduate
Research Fellowship Program.
The views and conclusions expressed are those of the authors and should
not be interpreted as representing the official views or policies of the
Army Research Office, the Department of Defense, or the U.S. Government.
\end{acks}

\bibliographystyle{ACM-Reference-Format}
\bibliography{main}

\end{document}